\pdfoutput=1
\documentclass[a4paper,12pt]{article}
\usepackage{amsmath}
\usepackage{amssymb}
\usepackage{amsfonts}
\usepackage{graphicx}
\usepackage{hhline}
\usepackage{xfrac}
\usepackage{bbold}

\usepackage[utf8]{inputenc}

\usepackage[square,numbers,comma,sort&compress]{natbib}
\usepackage[colorlinks=true,citecolor=blue,linkcolor=purple,urlcolor=purple]{hyperref}
\usepackage{geometry}
\usepackage[usenames,dvipsnames,svgnames,table]{xcolor}
\usepackage{booktabs}
\usepackage{slashed}

\def\varabstract{ }
\def\varkeywords{ }
\def\vararxivnumber{ }
\def\vartitle{ }
\def\varsubtitle{ }
\renewcommand{\title}[1]{\gdef\vartitle{#1}}

\renewcommand{\abstract}[1]{\gdef\varabstract{#1}}
\newcommand{\keywords}[1]{\gdef\varkeywords{#1}}

\newtoks\authtoks
\renewcommand{\author}[2][]{%
	\authtoks=\expandafter{\the\authtoks#2$^{#1}$\ }%
}
\newtoks\affiltoks
\newcommand{\affiliation}[2][]{%
    \affiltoks=\expandafter{\the\affiltoks{\item[$^{#1}$]#2}}%
}
\newtoks\emailtoks\newcounter{emailcounter}%
\newcommand{\emailAdd}[1]{%
\ifnum\theemailcounter>0\emailtoks=\expandafter{\the\emailtoks, \typeemail{#1}}%
\else\emailtoks=\expandafter{\typeemail{#1}}%
\fi
\stepcounter{emailcounter}}
\newcommand{\typeemail}[1]{\href{mailto:#1}{\tt #1}}

\renewcommand\maketitle{
	\newgeometry{margin=2cm}
	\pagestyle{empty}\setcounter{page}{0}
	{\huge\flushleft\sffamily\bfseries\vartitle\\\Large\varsubtitle\par}
\vskip6ex
{\large\bfseries\raggedright\sffamily\the\authtoks\par}
\vskip2ex
\begin{list}{}{%
\setlength{\leftmargin}{0.28cm}%
\setlength{\labelsep}{0pt}%
\setlength{\itemsep}{-3pt}%
\setlength{\topsep}{-\parskip}}
\itshape\small%
\the\affiltoks
\end{list}
\vskip2ex
\noindent\hspace{0.28cm}\begin{minipage}[l]{.9\textwidth}
\begin{flushleft}
\textit{E-mail:} \the\emailtoks
\end{flushleft}
\end{minipage}
\vskip5ex
\noindent{\renewcommand\baselinestretch{.9}\textsc{Abstract:}}\ \varabstract
\vskip5ex
\if!\varkeywords!\else\noindent{\textsc{Keywords:}}\ \varkeywords \vskip2ex\fi
\newpage
\restoregeometry
\pagestyle{plain}

\setcounter{footnote}{0}
}

\usepackage[square,numbers,comma,sort&compress]{natbib}

\definecolor{MS}{rgb}{0,0,1}

\usepackage{ifthen}

	\newcommand{\barlimc}[7]{
  \pgfmathparse{\mypos+0.3}
  \edef\mypos{\pgfmathresult}
		\node[left,scale=0.6] at (0,\mypos) {#1};
		\pgfmathparse{#3 > 5 ? 1 : 0}
		\ifthenelse{\pgfmathresult=1}{
			\fill[#2] ($(0,\mypos)+(0,-0.1)$) rectangle +(5,0.2);
			\fill[white] ($(0,\mypos)+(3.5,-0.1)$) rectangle +(0.3,0.2);
			\draw[decoration={zigzag},decorate,#2,very thick] (3.4,\mypos) to +(0.5,0);
			\node[left,scale=0.6] at (5,\mypos) {#3};
			}{
			\fill[#2] ($(0,\mypos)+(0,-0.1)$) rectangle +(#3,0.2);
			\node[left,scale=0.6] at (#3,\mypos) {#3};
		}
		\fill[#4] ($(0,\mypos)+(0,-0.1)$) rectangle +(#5,0.2);
		\node[left,scale=0.6] at (#5,\mypos) {#5};
		\fill[#6] ($(0,\mypos)+(0,-0.1)$) rectangle +(#7,0.2);
		\pgfmathparse{#7 <0.3 ? 1 : 0}
		\ifthenelse{\pgfmathresult=1}{
			\node[right,scale=0.6] at (0,\mypos) {#7};
		}{
		\node[left,scale=0.6] at (#7,\mypos) {#7};
	}
}

\title{\hspace{\fill}\mbox{\footnotesize\rm NCTS-PH/1904}%
\bigskip\bigskip\\
Lepton-flavor-violating semileptonic \boldmath$\tau$ decay and $K\to\pi\nu\bar\nu$}

\author[1,2,3]{Xiao-Gang He,}\emailAdd{hexg@phys.ntu.edu.tw}
\author[2,3]{ Jusak Tandean}\emailAdd{jtandean@phys.ntu.edu.tw}
\author[4]{ and German Valencia}\emailAdd{german.valencia@monash.edu}

\affiliation[1]{Tsung-Dao Lee Institute $\&$ School of Physics and Astronomy, SKLPPC,
Shanghai Jiao Tong University,
Shanghai 200240, China}
\affiliation[2]{Department of Physics, National Taiwan University,
Taipei 10617, Taiwan}
\affiliation[3]{Physics Division, National Center for Theoretical Sciences,
Hsinchu 30013, Taiwan}
\affiliation[4]{School of Physics and Astronomy, Monash University, Melbourne VIC-3800, Australia}

\abstract{We consider lepton-flavor violation in strangeness changing ($|\Delta S|=1$) semileptonic $\tau$-lepton decays arising from new physics encoded in a standard model effective Lagrangian. Its invariance under the standard model gauge group entails the relevance of other processes which can serve as complementary probes of the new physics operators. We show in particular that for some of them the bounds implied by current data on the rare kaon decays involving a neutrino pair, $K\to \pi\nu\bar\nu$, are stronger than the existing limits from direct searches for lepton-flavor-violating semileptonic $\tau$ decays. We discuss additional processes affected by the same operators and find that certain leptonic charged-meson decays also provide stricter constraints on a few more of them. Upcoming results of ongoing experiments such as Belle II and NA62 will further test the new physics parameter space.}


\begin{document}
\baselineskip=17pt \parskip=5pt

\maketitle

{\hypersetup{linkcolor=black}
  \tableofcontents}

\newpage

Interactions manifesting lepton-flavor violation (LFV) do not occur in the standard model (SM) with zero neutrino mass but are relatively common in new physics (NP) scenarios. There is a renewed interest in studying LFV for both theoretical and experimental reasons. On the theoretical side, the so-called `$B$-physics anomalies' constitute suggestive evidence for lepton-flavor universality violation \cite{Cerri:2018ypt}. Model building to account for them often gives rise to LFV as well. On the experimental side, there are a number of ongoing and forthcoming efforts that will improve upon the existing limits on LFV. Amongst them are LHCb~\cite{Cerri:2018ypt}, BESIII~\cite{Li:2012vk}, Belle II~\cite{Kou:2018nap}, and COMET~\cite{Calibbi:2017uvl}.

In a recent paper \cite{He:2019xxp} we have investigated the case of LFV in strangeness-changing ($|\Delta S|=1$) hyperon and kaon decays, where an initial strange (anti)quark decays. In the present paper we turn our attention to $\tau$-lepton decay where the strange quark (or antiquark) appears in the final state along with a down antiquark (or quark) and an electron or muon.
This kind of semileptonic $\tau$ transition has been addressed extensively in the past \cite{Valencia:1994cj,Ilakovac:1995km,Ilakovac:1995wc,Black:2002wh,Saha:2002kt,Kanemura:2005hr,Carpentier:2010ue,Dorsner:2011ai,Pich:2013lsa,Cai:2015poa}, besides its strangeness-conserving counterpart~\cite{Ilakovac:1995km,Ilakovac:1995wc,Gabrielli:2000te,Black:2002wh,Saha:2002kt,Kanemura:2005hr,Arganda:2008jj,He:2009tf,Arhrib:2009xf,Goto:2010sn,Carpentier:2010ue,Dorsner:2011ai,Daub:2012mu,Petrov:2013vka,Celis:2013xja,Pich:2013lsa,Hua:2014yna,Celis:2014asa,Cai:2015poa,Lami:2016vrs,Cai:2018cog}, under various NP contexts.

Following our earlier work \cite{He:2019xxp}, here we adopt a model-independent approach that starts from the most general effective Lagrangian involving dimension-six operators which respect the SM gauge symmetry and can generate \,$|\Delta S|=1$\, tau-flavor-violating interactions.
The resulting operators also contribute to other processes, in particular to ones where the lepton flavor is carried by a neutrino.
We explore the impact of these operators at tree level on various low-energy processes and map the constraints that can be extracted from the data available at the moment. We find that the so-called golden rare kaon decays, \,$K\to\pi\nu\bar\nu$,\, impose bounds on a number of the operators that are stricter by up to two orders of magnitude than the limits from direct searches for LFV in semileptonic $\tau$ decay.
Likewise, certain leptonic charged-meson decays also provide stronger restrictions on a few more of the operators.
The KOTO and NA62 experiments~\cite{Lurkin:2018gdo} can further tighten the constraints from \,$K\to\pi\nu\bar\nu$\, in the near future, and \mbox{Belle II} after achieving an integrated luminosity of 50 ab$^{-1}$ may improve upon the current limits on $\tau$ couplings exhibiting LFV by as much as an order of magnitude \cite{Kou:2018nap}.

\section{Effective Lagrangian\label{Leff}}

The most general effective Lagrangian constructed from SM fields, including an elementary Higgs, and invariant under the SM gauge group exists in the literature \cite{Buchmuller:1985jz,Grzadkowski:2010es}. The part of this Lagrangian containing the operators ${\cal Q}_k^{}$ pertinent to our discussion can be written as
\begin{equation} \label{Lnpo}
{\cal L}_{\textsc{np}}^{} \,=\, \frac{1}{\Lambda_{\textsc{np}}^2} \Bigg[
\raisebox{1pt}{\footnotesize$\displaystyle\sum_{\mbox{\scriptsize$k=1$}}^{\mbox{\scriptsize5}}$}\, {\cal C}_k^{ijxy} {\cal Q}_k^{ijxy} \,+\, \big( {\cal C}_6^{ijxy} {\cal Q}_6^{ijxy} + {\rm H.c.} \big) \Bigg] , ~~
\end{equation}
where $\Lambda_{\textsc{np}}$ stands for a heavy mass scale associated with the NP interactions, the coefficients ${\cal C}_{1,...,6}^{ijxy}$ are in general complex, and the family indices \,$i,j,x,y=1,2,3$\, are implicitly summed over. Explicitly,
\begin{align} \label{Qset}
{\cal Q}_1^{ijxy} & \,=\, \overline{q_i^{}} \gamma^\eta q_j^{}\, \overline{l_x^{}} \gamma_\eta^{} l_y^{} \,, & {\cal Q}_2^{ijxy} & \,=\, \overline{q_i^{}} \gamma^\eta \tau_{\texttt I}^{} q_j^{}\,
\overline{l_x^{}} \gamma_\eta^{} \tau_{\texttt I}^{} l_y^{} \,, & {\cal Q}_3^{ijxy} & \,=\,
\overline{d_i^{}}\gamma^\eta d_j^{}\, \overline{e_x^{}}\gamma_\eta^{} e_y^{} \,,
\nonumber \\
{\cal Q}_4^{ijxy} & \,=\, \overline{d_i^{}}\gamma^\eta d_j^{}\, \overline{l_x^{}}\gamma_\eta^{}
l_y^{} \,, & {\cal Q}_5^{ijxy} & \,=\, \overline{q_i^{}} \gamma^\eta q_j^{}\, \overline{e_x^{}}
\gamma_\eta^{} e_y^{} \,, &
{\cal Q}_6^{ijxy} & \,=\, \overline{l_i^{}} e_j^{}\, \overline{d_x^{}} q_y^{} \,.&
\end{align}
The notation is standard and detailed in Ref.\,\cite{He:2019xxp}. For convenience, we work in the mass basis of the down-type fermions, where
\begin{align} \label{ql}
q_i^{} & \,=\, P_L^{} \left(\!\begin{array}{c} \raisebox{1pt}{\footnotesize$\sum$}_j^{~}
\big({\cal V}_{\textsc{ckm}}^\dagger\big)_{\!ij} \textsl{\texttt U}_j^{} \\ \textsl{\texttt D}_i^{} \end{array}\!\right) , &
l_i^{} & \,=\, P_L^{} \left(\!\begin{array}{c} \raisebox{1pt}{\footnotesize$\sum$}_j^{}\,
({\cal U}_{\textsc{pmns}})_{ij}^{}\nu_j^{} \\ \textsl{\texttt E}_i^{} \end{array}\!\right) , &
e_i^{} & \,=\, P_R^{} \textsl{\texttt E}_i^{} \,, & d_i^{} & \,=\, P_R^{} \textsl{\texttt D}_i^{} \,,
\end{align}
with \,${\cal V}_{\textsc{ckm}}$\, $({\cal U}_{\textsc{pmns}})$ being the Cabibbo-Kobayashi-Maskawa quark (Pontecorvo-Maki-Nakagawa-Sakata neutrino) mixing matrix. All the fields appearing in Eq.\,(\ref{ql}) are thus mass eigenstates.
The part of ${\cal L}_{\textsc{np}}$ containing the operators responsible for \,$|\Delta S|=1$\, semileptonic \,$\tau^\pm\to\ell^\pm$\, transitions, with \,$\ell=\textsl{\texttt E}_{1,2}=e,\mu$,\, can then be expressed as
\begin{equation} \label{Lnp}
{\cal L}_{\textsc{np}}^{} \,\supset\, \frac{1}{\Lambda_{\textsc{np}}^2}\,
\raisebox{1pt}{\footnotesize$\displaystyle\sum_{\mbox{\scriptsize$k$=1}}^{\mbox{\scriptsize6,6$\prime$}}$}~
\raisebox{1pt}{\footnotesize$\displaystyle\sum_{\mbox{\scriptsize$n$=1}}^{\scriptsize2}$}
\Big( c_k^{\textsl{\texttt E}_n^{}\tau} Q_k^{\textsl{\texttt E}_n^{}\tau}+c_k^{\tau\textsl{\texttt E}_n^{}} Q_k^{\tau\textsl{\texttt E}_n^{}} \Big)
\,+\, {\rm H.c.} \,,
\end{equation}
where
\,$c_{\hat\kappa}^{\textsl{\texttt E}_n^{}\tau(\tau\textsl{\texttt E}_n^{})} = {\cal C}_{\hat\kappa}^{12n3(123n)}$\, and
\,$Q_{\hat\kappa}^{\textsl{\texttt E}_n^{}\tau(\tau\textsl{\texttt E}_n^{})} = {\cal Q}_{\hat\kappa}^{12n3(123n)}$\, for
\,$\hat\kappa=1,...,5$,\,
$c_6^{\textsl{\texttt E}_n^{}\tau(\tau\textsl{\texttt E}_n^{})} = {\cal C}_6^{n312(3n12)}$,\,
$Q_6^{\textsl{\texttt E}_n^{}\tau(\tau\textsl{\texttt E}_n^{})} = {\cal Q}_6^{n312(3n12)}$,\,
$c_{6\prime}^{\textsl{\texttt E}_n^{}\tau(\tau\textsl{\texttt E}_n^{})} = {\cal C}_6^{3n21(n321)*}$,\, and
\,$Q_{6\prime}^{\textsl{\texttt E}_n^{}\tau(\tau\textsl{\texttt E}_n^{})} =
{\cal Q}_6^{3n21(n321)\dagger}$.\,

\section{Amplitudes and rates\label{rates}}

\subsection{\bf\boldmath$|\Delta S|=1$ semileptonic $\tau$ decays\label{rates}}

We treat first $\tau$ decay into a charged lepton $\ell$ plus a pseudoscalar meson $P$ or a vector meson~$V$, on which there are direct search data.
For \,$\tau^-\to\ell^-P$\, and \,$\tau^-\to\ell^-V$\, the amplitudes have the general forms, respectively,
\begin{align} \label{tau->lP}
{\cal M}_{\tau\to\ell P}^{} & \,=\, i \bar u_\ell^{} \Big( {\cal S}_P^\ell
\,+\, \gamma_5^{}\, {\cal P}_P^\ell \Big) u_\tau^{} \,,
\\ \label{tau->lV}
{\cal M}_{\tau\to\ell V}^{} & \,=\, \bar u_\ell^{}\, \slashed\varepsilon_V^{}
\Big( {\cal V}_V^\ell \,+\, \gamma_5^{}\, {\cal A}_V^\ell \Big) u_\tau^{} \,,
\end{align}
which lead to the decay rates
\begin{align} \label{Gtau->lP}
\Gamma_{\tau\to\ell P}^{} & \,=\, \frac{{\cal K}^{1/2}\big(m_\tau^2,m_\ell^2,m_P^2\big)}
{16\pi\, m_\tau^3} \Big\{ \big[ (m_\tau^{}+m_\ell^{})^2-m_P^2\big] \big|{\cal S}_P^\ell\big| \raisebox{2pt}{$^2$}
+ \big[ (m_\tau^{}-m_\ell^{})^2-m_P^2\big] \big|{\cal P}_P^\ell\big| \raisebox{2pt}{$^2$} \Big\} \,,
\nonumber \\
\Gamma_{\tau\to\ell V}^{} & \,=\, \frac{{\cal K}^{1/2}\big(m_\tau^2,m_\ell^2,m_V^2\big)}
{16\pi\, m_\tau^3\, m_V^2} \begin{array}[t]{l} \!\! \Big\{
\big[ \tilde{\textsc k}\big(m_\tau^2,m_\ell^2,m_V^2\big)-6 m_\tau^{}m_\ell^{}m_V^2 \big]
\big|{\cal V}_V^\ell\big| \raisebox{2pt}{$^2$}
\\ +\;
\big[ \tilde{\textsc k}\big(m_\tau^2,m_\ell^2,m_V^2\big)+6 m_\tau^{}m_\ell^{}m_V^2 \big]
\big|{\cal A}_V^\ell\big| \raisebox{2pt}{$^2$} \Big\} \,, \end{array}
\end{align}
where \,${\cal K}(x,y,z)=(x-y-z)^2-4y z$\, and \,$\tilde{\textsc k}(x,y,z)=(x-y)^2+(x+y)z-2z^2$.\,

For \,$\tau^-\to\ell^-K_S$\, the hadronic matrix elements which do not vanish are
\begin{align}
\big\langle K^0\big|\overline{d}\gamma^\eta\gamma_5^{}s\big|0\big\rangle & =
\big\langle\,\overline{\!K}{}^0\big|\overline{s}\gamma^\eta\gamma_5^{}d\big|0\big\rangle
=\, i f_K^{}p_K^\eta \,, & ~
\big\langle K^0\big|\overline{d}\gamma_5^{}s\big|0\big\rangle & =
\big\langle\,\overline{\!K}{}^0\big|\overline{s}\gamma_5^{}d\big|0\big\rangle =\,
i B_0^{} f_K^{} \,, ~
\end{align}
where $f_K^{}$ is the kaon decay constant and \,$B_0^{}=m_{K^0}^2/(m_d^{}+m_s^{})$.\,
Applying them to the operators in Eq.\,(\ref{Lnp}) for the amplitude in Eq.\,(\ref{tau->lP}), with the approximation \,$\sqrt2\, K_S=K^0-\overline K{}^0$\, we get
\begin{align} \label{lK}
{\cal S}_{K_S}^\ell & \,=\, f_K^{}\, \frac{(m_\tau^{}-m_\ell^{}) \big( \tilde{\textsc v}_{\ell\tau}^{} - \tilde{\textsc v}_{\tau\ell}^* \big) + B_{0\,}^{} \big( \tilde{\textsc s}_{\ell\tau}^{} + \tilde{\textsc s}_{\tau\ell}^* \big)}{4\sqrt2\,\Lambda_{\textsc{np}}^2} \,,
\nonumber \\
{\cal P}_{K_S}^\ell & \,=\, f_K^{}\, \frac{(m_\tau^{}+m_\ell^{}) \bigl(-\tilde{\textsc a}_{\ell\tau}^{} + \tilde{\textsc a}_{\tau\ell}^*\bigr) + B_{0\,}^{} \big( \tilde{\textsc p}_{\ell\tau}^{} - \tilde{\textsc p}_{\tau\ell}^* \big)}{4\sqrt2\,\Lambda_{\textsc{np}}^2} \,,
\end{align}
where
\begin{align}
\tilde{\textsc v}_{\textsl{\texttt X}\textsl{\texttt Y}}^{} & \,=\,  c_1^{\textsl{\texttt X}\textsl{\texttt Y}} + c_2^{\textsl{\texttt X}\textsl{\texttt Y}} - c_3^{\textsl{\texttt X}\textsl{\texttt Y}} - c_4^{\textsl{\texttt X}\textsl{\texttt Y}} + c_5^{\textsl{\texttt X}\textsl{\texttt Y}} \,, &
\tilde{\textsc a}_{\textsl{\texttt X}\textsl{\texttt Y}}^{} & \,=\, -c_1^{\textsl{\texttt X}\textsl{\texttt Y}} - c_2^{\textsl{\texttt X}\textsl{\texttt Y}} - c_3^{\textsl{\texttt X}\textsl{\texttt Y}} + c_4^{\textsl{\texttt X}\textsl{\texttt Y}} + c_5^{\textsl{\texttt X}\textsl{\texttt Y}} \,, &
\nonumber \\ \label{sp}
\tilde{\textsc s}_{\textsl{\texttt X}\textsl{\texttt Y}}^{} & \,=\, c_6^{\textsl{\texttt X}\textsl{\texttt Y}} - c_{6\prime}^{\textsl{\texttt X}\textsl{\texttt Y}} \,, &
\tilde{\textsc p}_{\textsl{\texttt X}\textsl{\texttt Y}}^{} & \,=\, c_6^{\textsl{\texttt X}\textsl{\texttt Y}} + c_{6\prime}^{\textsl{\texttt X}\textsl{\texttt Y}} \,.
\end{align}
Similarly, for \,$\tau^-\to\ell^-K^{*0},\ell^-\,\overline{\!K}{}^{*0}$\, the nonzero mesonic matrix elements are
\begin{align}
\big\langle K^{*0}\big|\overline{d}\gamma^\eta s\big|0\big\rangle & \,=\,
\big\langle\,\overline{\!K}{}^{*0}\big|\overline{s}\gamma^\eta d\big|0\big\rangle  \,=\,
\varepsilon_{K^*\,}^\eta f_{K^*\,}^{} m_{K^*}^{} \,, &
\end{align}
where $\varepsilon_{K^*}$ and $f_{K^*}^{}$ are, respectively, the polarization vector and decay constant of $K^*$.
Hence from Eqs.\,\,(\ref{Lnp}) and (\ref{tau->lV}) follow
\begin{align} \label{lK*}
{\cal V}_{K^{*0}}^\ell & \,=\,
\frac{f_{K^*\,}^{} m_{K^*}^{}}{4\Lambda_{\textsc{np}}^2}\, \texttt V_{\ell\tau}^{} \,, &
{\cal A}_{K^{*0}}^\ell & \,=\,
\frac{f_{K^*\,}^{} m_{K^*}^{}}{4\Lambda_{\textsc{np}}^2}\, \texttt A_{\ell\tau}^{} \,,
\nonumber \\
{\cal V}_{\,\overline{\!K}{}^{*0}}^\ell & \,=\,
\frac{f_{K^*\,}^{} m_{K^*}^{}}{4\Lambda_{\textsc{np}}^2}\, \texttt V_{\tau\ell}^* \,, &
{\cal A}_{\,\overline{\!K}{}^{*0}}^\ell & \,=\,
\frac{f_{K^*\,}^{} m_{K^*}^{}}{4\Lambda_{\textsc{np}}^2}\, \texttt A_{\tau\ell}^* &
\end{align}
with
\begin{align}
\texttt V_{\textsl{\texttt X}\textsl{\texttt Y}}^{} & \,=\, -c_1^{\textsl{\texttt X}\textsl{\texttt Y}} - c_2^{\textsl{\texttt X}\textsl{\texttt Y}} - c_3^{\textsl{\texttt X}\textsl{\texttt Y}} - c_4^{\textsl{\texttt X}\textsl{\texttt Y}} - c_5^{\textsl{\texttt X}\textsl{\texttt Y}} \,, &
\texttt A_{\textsl{\texttt X}\textsl{\texttt Y}}^{} & \,=\,  c_1^{\textsl{\texttt X}\textsl{\texttt Y}} + c_2^{\textsl{\texttt X}\textsl{\texttt Y}} - c_3^{\textsl{\texttt X}\textsl{\texttt Y}} + c_4^{\textsl{\texttt X}\textsl{\texttt Y}} - c_5^{\textsl{\texttt X}\textsl{\texttt Y}} \,. &
\end{align}
Clearly, \,$\tau^\pm\to\ell^\pm K^{*0},\ell^\pm\,\overline{\!K}{}^{*0}$\, can access only the coupling combinations $\texttt V_{\ell\tau,\tau\ell}$ and $\texttt A_{\ell\tau,\tau\ell}$, while \,$\tau^\pm\to\ell^\pm K_S$\, cannot probe them.

We turn now to the three-body modes \,$\tau^-\to\ell^-\pi^\pm K^\mp$,\, on which empirical information also exists.
The relevant hadronic matrix elements in this case are given by
\begin{align}
\big\langle\pi^-K^+\big|\overline{d}\gamma^\eta s\big|0\big\rangle & \,=\,
-\big\langle\pi^+K^-\big|\overline{s}\gamma^\eta d\big|0\big\rangle \,=\,
f_+^{}\, \big(p_\pi^\eta - p_K^\eta\big) - f_-^{}\, \hat q^\eta \,,
\nonumber \\
\big\langle\pi^-K^+\big|\overline{d} s\big|0\big\rangle & \,=\,
\big\langle\pi^+K^-\big|\overline{s} d\big|0\big\rangle \,=\, \tilde B_0^{}\, f_0^{} \,, \hspace{4em}
f_-^{} \,=\, \big(f_0^{}-f_+^{}\big) \frac{\Delta_{K\pi}^2}{\hat q^2} \,, &
\end{align}
where $f_+^{}$ and $f_0^{}$ denote form factors which are functions of $\hat q^2$,
\begin{align}
\hat q & \,=\, p_\pi^{} +  p_K^{} \,, & \Delta_{K\pi}^2 & \,=\, m_{K^+}^2 - m_{\pi^+}^2 \,, &
\tilde B_0^{} & \,=\, \frac{\Delta_{K\pi}^2}{m_s^{}-m_d^{}} \,. &
\end{align}
Accordingly, the amplitude for \,$\tau^-\to\ell^-\pi^-K^+$\, is
\begin{align}
{\cal M}_{\tau\to\ell\pi^-K^+}^{} & \,=\, \bar u_\ell^{} \Big( {\cal S}_{\pi^-K^+}^\ell
+ \gamma_5^{}\,{\cal P}_{\pi^-K^+}^\ell \Big) u_\tau^{} \,, &
\end{align}
where
\begin{align}
{\cal S}_{\pi^-K^+}^\ell & \,=\, \bigl[ - 2 f_+^{}\, \slashed p{}_K^{} + \big(f_+^{}-f_-^{}\big)
(m_\tau^{}-m_\ell) \big] \frac{\texttt V_{\ell\tau}}{4\Lambda_{\textsc{np}}^2}
+ \frac{\tilde B_0^{}\, f_0^{}\, \texttt S_{\ell\tau}^{}}{4\Lambda_{\textsc{np}}^2} \,, &
\\
{\cal P}_{\pi^-K^+}^\ell  & \,=\, \big[ 2 f_+^{}\, \slashed p{}_K^{} - \big(f_+^{}-f_-^{}\big)
(m_\tau^{}+m_\ell) \big] \frac{\texttt A_{\ell\tau}^{}}{4\Lambda_{\textsc{np}}^2}
+ \frac{\tilde B_0^{}\, f_0^{}\, \texttt P_{\ell\tau}^{}}{4\Lambda_{\textsc{np}}^2} \,, &
\end{align}
with
\,$\texttt S_{\ell\tau}^{}=-c_6^{\ell\tau}-c_{6\prime}^{\ell\tau}=-\tilde{\textsc p}_{\ell\tau}^{}$\, and
\,$\texttt P_{\ell\tau}^{}=-c_6^{\ell\tau}+c_{6\prime}^{\ell\tau}=-\tilde{\textsc s}_{\ell\tau}^{}$.\,
Its differential rate is then
\begin{align} \label{G't2lpiK}
\frac{d\Gamma_{\tau\to\ell\pi^-K^+}^{}}{d\hat s} \,=\, \frac{\lambda_{\tau\ell}^{1/2\,}
\lambda_{\pi^+K^+}^{1/2\,} \big|f_0^{}\big|\raisebox{1pt}{$^2$}}{256\pi^3\, m_\tau^3\,
\Lambda_{\textsc{np}}^4} & \Bigg\{ \Bigg[ \lambda_{\pi^+K^+\,}^{}
\big|f_+^{}\big|\raisebox{1pt}{$^2$}\; \frac{\lambda_{\tau\ell}^{}
+ 3 \hat\sigma_-^{}\hat s}{3 \big|f_0^{}\big|\raisebox{1pt}{$^2$}\, \hat s^3} + \Delta_{K\pi}^4\,
\frac{\lambda_{\tau\ell}^{}+\hat\sigma_+^{}\hat s}{\hat s^3} \Bigg] \frac{|\texttt V_{\ell\tau}|^2}{16}
\nonumber \\ & +\,
\Bigg[ \lambda_{\pi^+K^+\,}^{} \big|f_+^{}\big|\raisebox{1pt}{$^2$}\;
\frac{\lambda_{\tau\ell}^{}+3\hat\sigma_+^{}\hat s}
{3 \big|f_0^{}\big|\raisebox{1pt}{$^2$}\, \hat s^3} + \Delta_{K\pi}^4\, \frac{\lambda_{\tau\ell}^{}
+ \hat\sigma_-^{}\hat s}{\hat s^3} \Bigg] \frac{|\texttt A_{\ell\tau}|^2}{16}
\nonumber \\ & +\,
\frac{\Delta_{K\pi\,}^2 \tilde B_0^{}}{8 \hat s^2}\,
{\rm Re} \big( \hat\mu_+^{}\, \hat\sigma_-^{}\, \texttt A_{\ell\tau\,}^* \texttt P_{\ell\tau}^{}
- \hat\mu_-^{}\, \hat\sigma_+^{}\, \texttt V_{\ell\tau\,}^* \texttt S_{\ell\tau}^{} \big)
\nonumber \\ & +\,
\frac{\tilde B_0^2}{16 \hat s} \big( \hat\sigma_+^{}\, |\texttt S_{\ell\tau}|^2
+ \hat\sigma_-^{}\, |\texttt P_{\ell\tau}|^2 \big) \Bigg\} \,, &
\end{align}
where
\begin{align}
\hat s & \,=\, \hat q^2 \,, & \lambda_{XY}^{} & \,=\, {\cal K}\big(m_X^2,m_Y^2,\hat s\big) \,, &
\hat\sigma_\pm^{} & \,=\, \hat\mu_\pm^2-\hat s \,, & \hat\mu_\pm^{} & \,=\, m_\tau\pm m_\ell^{} \,. &
\end{align}
The differential rate of \,$\tau^-\to\ell^-\pi^+K^-$\, is also given by Eq.\,(\ref{G't2lpiK}) but with \,$(\texttt V_{\ell\tau},\texttt A_{\ell\tau},\texttt S_{\ell\tau},\texttt P_{\ell\tau})$\, changed to \,$(\texttt V_{\tau\ell},\texttt A_{\tau\ell},-\texttt S_{\tau\ell},\texttt P_{\tau\ell})$.\,
We observe that, unlike  \,$\tau\to\ell K^{*0},\ell\,\overline{\!K}{}^{*0}$,\, these three-body modes are sensitive to all the operators with parity-even quark parts.

\subsection{\bf Other modes\label{others}}

The required SU(2)$_L$-gauge-invariance of $Q_k^{\ell\tau}$ and $Q_k^{\tau\ell}$ in Eq.\,(\ref{Lnp}) implies that some of these operators involve left-handed quark and/or lepton doublets and therefore can influence additional processes.
The related couplings are summarized in Appendix \ref{feynrules} and can generate transitions with one or two neutrinos.
Here we discuss the extra modes which may offer complementary restrictions on the couplings.

\noindent
{\, $\scriptscriptstyle\blacksquare$ ~\boldmath$K^+\to\pi^+\nu\bar\nu$\, and \,$K_L\to\pi^0\nu\bar\nu$}

The operators in Eq.\,(\ref{Lnp}) with a pair of left-handed lepton doublets provide the $(\bar ds)(\bar\nu_\ell\nu_\tau)$ interaction listed in Table \ref{tabFR}, as well as its $\bar\nu_\tau\nu_\ell$ counterpart.
Since they have neutrino-flavor combinations that are different from those in the SM amplitudes, the former have no interference with latter and cause the \,$K\to\pi\nu\bar\nu$\, rates to exceed their SM values, as the neutrinos are not detected.
The resulting modifications $\Delta{\cal B}_{K^+}$ and $\Delta{\cal B}_{K_L}$ to the SM branching fractions can be inferred from Eqs.\,(9)-(10) in Ref.\,\cite{He:2018uey} to be
\begin{align} \label{K2pnn}
\Delta{\cal B}_{K^+}^{} & \,=\, {\mathcal B}(K^+\to\pi^+\nu\nu)_{\textsc{np}}^{} \,=\,
\frac{\tilde\kappa_+^{}}{3}~ \raisebox{4pt}{\footnotesize$\displaystyle\sum_{\mbox{\scriptsize$\ell=e,\mu$}}$}\, \big( |W_{\ell\tau}|^2 + |W_{\tau\ell}|^2 \big) , &
\nonumber \\
\Delta{\cal B}_{K_L}^{} & \,=\, {\mathcal B}(K_L\to\pi^0\nu\bar\nu)_{\textsc{np}}^{} \,=\,
\frac{\kappa_L^{}}{12}~ \raisebox{4pt}{\footnotesize$\displaystyle\sum_{\mbox{\scriptsize$\ell=e,\mu$}}$}\, \big|W_{\ell\tau}^{}-W_{\tau\ell}^*\big|^2 \,, &
\\ \label{Wlt}
W_{\textsl{\texttt X}\textsl{\texttt Y}}^{} & \,\simeq\, 9700 \bigg(\frac{1\rm~TeV}{\Lambda_{\textsc{np}}}\bigg)\raisebox{8pt}{$\!^2$} \big( c_1^{\textsl{\texttt X}\textsl{\texttt Y}} -c_2^{\textsl{\texttt X}\textsl{\texttt Y}} + c_4^{\textsl{\texttt X}\textsl{\texttt Y}} \big) \,, &
\end{align}
where \cite{Buras:2004uu} \,$\tilde\kappa_+^{} = 5.17\times10^{-11}$\, and \,$\kappa_L^{}=2.23\times 10^{-10}$.\,

\medskip

\noindent
{\, $\scriptscriptstyle\blacksquare$ ~\boldmath$\tau^-\to\ell^-\pi^0$ and $\tau^-\to\ell^-\rho^0$}

These are induced by the $(\bar u u) \big(\bar\ell\tau\big)$ couplings in Table \ref{tabFR} from the operators with a pair of left-handed quark doublets.
The pertinent mesonic matrix elements are
\,$\big\langle\pi^0\big|\overline{u}\gamma^\eta\gamma_5^{}u\big|0\big\rangle =
i f_\pi^{}p_\pi^\eta/\sqrt2$\, and
\,$\big\langle\rho^0\big|\overline{u}\gamma^\eta u\big|0\big\rangle =
\varepsilon_\rho^\eta f_\rho^{} m_\rho^{}/\sqrt2$,\,
where $f_{\pi(\rho)}^{}$ is the pion ($\rho$ meson) decay constant and $\varepsilon_\rho^{}$ is the $\rho$ polarization vector.
The \,$\tau\to\ell\pi,\ell\rho$\, rates have, respectively, the forms in Eq.\,(\ref{Gtau->lP}) with
\begin{align} \label{lpi}
S_{\pi^0}^\ell & = \frac{i f_\pi^{} V_{ud}^{}V_{us}^*}{4\sqrt2\, \Lambda_{\textsc{np}}^2}
\Big(c_1^{\ell\tau}-c_2^{\ell\tau}+c_5^{\ell\tau}\Big) \big(m_\tau^{}-m_\ell^{}\big) \,, &
P_{\pi^0}^\ell & = \frac{i f_\pi^{} V_{ud}^{}V_{us}^*}{4\sqrt2\, \Lambda_{\textsc{np}}^2}
\Big(c_1^{\ell\tau}-c_2^{\ell\tau}-c_5^{\ell\tau}\Big) \big(m_\tau^{}+m_\ell^{}\big) \,,
\nonumber \\
{\cal V}_{\rho^0}^\ell & = \frac{-f_\rho^{} m_\rho^{} V_{ud}^{}V_{us}^*}
{4\sqrt2\, \Lambda_{\textsc{np}}^2} \Big(c_1^{\ell\tau}-c_2^{\ell\tau}+c_5^{\ell\tau}\Big) \,, &
{\cal V}_{\rho^0}^\ell & = \frac{f_\rho^{} m_\rho^{} V_{ud}^{}V_{us}^*}
{4\sqrt2\, \Lambda_{\textsc{np}}^2} \Big(c_1^{\ell\tau}-c_2^{\ell\tau}-c_5^{\ell\tau}\Big) \,.
\end{align}
These are suppressed by the CKM factor \,$|V_{ud}V_{us}|\simeq0.22$\, compared to their counterparts in Eqs.\,\,(\ref{lK}) and (\ref{lK*}).

\medskip

\noindent
{\, $\scriptscriptstyle\blacksquare$ ~\boldmath$J/\psi\to\ell^\mp\tau^\pm$}

Like the preceding case, Eq.\,(\ref{Lnp}) includes the $(\bar cc)\big(\bar\ell\tau\big)$ interaction, listed in Table \ref{tabFR}, which brings about the charmonium decay \,$J/\psi\to\ell^-\tau^+$\, and is also suppressed by \,$|V_{cd}V_{cs}|\simeq0.22$.\,
With $m_\ell^{}$ neglected, the rate of this mode is
\begin{align} \label{GJ2lt}
\Gamma_{J/\psi\to\ell^-\tau^+}^{} & \,=\,
\frac{f_{J/\psi}^2\, |V_{cd}V_{cs}|^2}{192\pi \Lambda_{\textsc{np}}^{\raisebox{1pt}{\scriptsize$4$}} m_{J/\psi}^3} \big(m_{J/\psi}^2-m_\tau^2\big)\raisebox{1pt}{$^2$}
\big(2 m_{J/\psi}^2+m_\tau^2\big) \Big( \big|c_1^{\ell\tau}-c_2^{\ell\tau}\big|\raisebox{2pt}{$^2$}
+ \big|c_5^{\ell\tau}\big|\raisebox{2pt}{$^2$} \Big) \,, &
\end{align}
where the $J/\psi$ decay constant $f_{J/\psi}^{}$ is defined by
\,$\langle0|\overline{c}\gamma^\kappa c|J/\psi\rangle=\varepsilon_{J/\psi\,}^\kappa
f_{J/\psi\,}^{} m_{J/\psi}^{}$,\,
which involves the $J/\psi$ polarization vector $\varepsilon_{J/\psi}$.
The rate of \,$J/\psi\to\ell^+\tau^-$\, equals $\Gamma_{J/\psi\to\ell^-\tau^+}$ but with $c_k^{\ell\tau}$ replaced by $c_k^{\tau\ell}$.

\medskip

\noindent
{\, $\scriptscriptstyle\blacksquare$ ~\boldmath${\cal P}^+\to\ell^+\nu$}

The couplings in the last four rows of Table \ref{tabFR} or the analogous couplings with $\ell$ and $\tau$ interchanged can affect the SM-dominated leptonic decay \,${\cal P}^+\to\ell^+\nu$\, of a charged pseudoscalar meson \,${\cal P}^+\sim\textsc u\bar{\textsc d}$,\, where \,$\textsc u=u,c$\, and \,$\textsc d=d,s$.\,
The biggest impact comes from the (pseudo)scalar operators, which are not subject to helicity suppression, with \,$\ell=e$,\, in which case the SM amplitude is the most helicity-suppressed.
With only $O_{6(\prime)}^{e\tau,\tau e}$ being present, we derive the modification $\Delta\Gamma_{{\cal P}^+\to e^+\nu}$ to the SM rate of \,${\cal P}^+\to e^+\nu$\, for \,${\cal P}=\pi,K,D,D_s$\, to be \cite{He:2019xxp}
\begin{align} \label{DGM2enu} & \hspace{5em}
\Delta\Gamma_{{\cal P}^+\to e^+\nu}^{} \,=\,
\frac{\big|\hat{\texttt C}_{\cal P}\big|^2 f_{\cal P}^2\, m_{\cal P}^5}{64 \pi \Lambda^4_{\textsc{np}\,}(m_{\textsc u}+m_{\textsc d})^2} \,,
\\ \label{CM}
\hat{\texttt C}_\pi^{} & \,=\, c_6^{\tau e\,} V_{us}^* \,, ~~~~~ ~~~~~
\hat{\texttt C}_K^{}     \,=\, c_{6\prime}^{e\tau*\,} V_{ud}^*  \,, ~~~~~ ~~~~~
\hat{\texttt C}_D^{}     \,=\, c_6^{\tau e\,} V_{cs}^* \,, ~~~~~ ~~~~~
\hat{\texttt C}_{D_s}^{} \,=\, c_{6\prime}^{e\tau*\,} V_{cd}^* \,,
\end{align}
with the $\cal P$ decay constant $f_{\cal P}^{}$ being defined by \,$\langle0|\overline{\textsc d}\gamma_5^{}\textsc u|{\cal P}^+\rangle=i f_{\cal P}^{}m_{\cal P}^2/(m_{\textsc u}+m_{\textsc d})$\, and the lepton masses ignored.
Note that there is no interference with the SM contribution as the neutrino is of the wrong flavor \cite{Valencia:1994cj}.

It is worth pointing out here that among the operators ${\cal Q}_{1,...,6}^{ijxy}$ in Eq.\,(\ref{Qset}) there are those not relevant to $ds\ell\tau$ interactions which can in general also influence some of the others listed in table\,\,\ref{tabFR}, particularly the ones involving up-type quarks.
For example, ${\cal Q}_1^{1113}$ in our mass basis, specified by Eq.\,(\ref{ql}), contributes to $(\bar u u,\bar u c)\big(\bar e\tau,\bar\nu_e \nu_\tau\big)$ couplings.\footnote{Moreover, there are operators \cite{Buchmuller:1985jz,Grzadkowski:2010es} not listed in Eq.\,(\ref{Qset}), such as \,$Q_{lu}^{ij13}=\overline{u_i^{}}\gamma^\eta u_j^{}\, \overline{l_1^{}}\gamma_\eta^{}l_3^{}$,\, which contribute to these same couplings.\medskip}
In dealing with the constraints from the preceding extra processes, we will ignore these other operators.
One may regard this as an additional model assumption, or basis dependence, implicit in our analysis.

\section{Numerical results\label{numeric}}

\subsection{\bf\boldmath$|\Delta S|=1$ semileptonic $\tau$ decays\label{tau2lds}}

We treat the two-body modes with the amplitude terms in Eqs.\,\,(\ref{lK}) and (\ref{lK*}) and the decay rates in~Eq.\,(\ref{Gtau->lP}).
The required decay constants are \,$f_K^{}=155.6(4)$\,MeV \cite{Tanabashi:2018oca} and \,$f_{K^*}^{}=206(6)$\,MeV,\, the latter having been extracted from the data on \,$\tau^-\to\nu K^{*-}$\, under the assumptions of isospin symmetry and no NP in this channel.\footnote{We have employed \,${\cal B}(\tau^-\to\nu K^{*-})=0.0120(7)$\, and \,$m_{K^{*-}}=895.5(8)$ MeV\, from \cite{Tanabashi:2018oca}.
For $V_{us}$ and the other CKM matrix elements needed in our numerical work, we adopt the results of \cite{Charles:2015gya} with the latest updates available at http://ckmfitter.in2p3.fr.\medskip}
In the calculation of rates in this section, we use the central values of hadron masses supplied by Ref.\,\cite{Tanabashi:2018oca} and, when occur, quark masses at a renormalization scale of 2\,GeV,\, namely \,$(m_u,m_d,m_s)=(2.2,4.7,95)$\,MeV\, \cite{Tanabashi:2018oca} and \,$m_c=1.1$\,GeV.\footnote{The latter has been rescaled from \,$m_c(m_c)=1.275$\,GeV\, \cite{Tanabashi:2018oca}.}

Thus, we arrive at the branching fractions
\begin{align} \label{Btau2eKS}
{\cal B}(\tau^-\to e^-K_S) & \,=\, 3.2 \begin{array}[t]{l} \!\! \Big[
\big| \tilde{\textsc v}_{e\tau}^{} - \tilde{\textsc v}_{\tau e}^* + 1.4\, \big(
\tilde{\textsc s}_{e\tau}^{} + \tilde{\textsc s}_{\tau e}^* \big) \big| \raisebox{2pt}{$^2$}
\\ +\; \displaystyle
\big| \tilde{\textsc a}_{e\tau}^{} - \tilde{\textsc a}_{\tau e}^* - 1.4\, \big(
\tilde{\textsc p}_{e\tau}^{} - \tilde{\textsc p}_{\tau e}^* \big) \big| \raisebox{2pt}{$^2$}
\Big] \frac{10^7\rm\,GeV^4}{\Lambda_{\textsc{np}}^4} \,, \end{array}
\nonumber \\
{\cal B}(\tau^-\to\mu^-K_S) & \,= \begin{array}[t]{l} \!\! \Big[ 3.2\, \big|
\tilde{\textsc v}_{\mu\tau}^{} - \tilde{\textsc v}_{\tau\mu}^* + 1.5\, \big(
\tilde{\textsc s}_{\mu\tau}^{} + \tilde{\textsc s}_{\tau\mu}^* \big) \big| \raisebox{2pt}{$^2$}
\\ +\; \displaystyle
3.1\, \big| \tilde{\textsc a}_{\mu\tau}^{} - \tilde{\textsc a}_{\mu\ell}^* - 1.3\, \big(
\tilde{\textsc p}_{\mu\tau}^{} - \tilde{\textsc p}_{\tau\mu}^* \big) \big| \raisebox{2pt}{$^2$} \Big]
\frac{10^7\rm\,GeV^4}{\Lambda_{\textsc{np}}^4} \,, \end{array}
\end{align}
\begin{align} \label{Btau2eK*}
{\cal B}\big(\tau^-\to e^-K^{*0}\big) & \,=\, 1.1\,\big(|\texttt V_{e\tau}|^2+|\texttt A_{e\tau}|^2\big)
\frac{10^8\rm\,GeV^4}{\Lambda_{\textsc{np}}^4} \,,   \vphantom{\int_\int^{\int^|}}
\nonumber\\
{\cal B}\big(\tau^-\to\mu^-K^{*0}\big) & \,=\, \big( 1.0\,|\texttt V_{\mu\tau}|^2
+ 1.2\,|\texttt A_{\mu\tau}|^2 \big) \frac{10^8\rm\,GeV^4}{\Lambda_{\textsc{np}}^4} \,, &
\end{align}
and those for \,$\tau^-\to\ell^-\,\overline{\!K}{}^{*0}$,\, which are the same as the ones in Eq.\,(\ref{Btau2eK*}) but with the subscript $\ell\tau$ changed to~$\tau\ell$.
To evaluate the three-body case with Eq.\,(\ref{G't2lpiK}), we need the $\pi K$ form-factors $f_+^{}$ and $f_0^{}$ as functions of~$\hat s$.
Assuming isospin symmetry, we adopt the $\pi^-K_S$ invariant-mass spectrum which has been extracted from the study of \,$\tau^-\to\nu\pi^-K_S$\, by the Belle Collaboration~\cite{Epifanov:2007rf}.\footnote{We have picked the $K_S^{}\pi$ mass spectrum in the \,$K_0^*(800)+K^*(892)+K^*(1410)$\, model determined by Belle~\cite{Epifanov:2007rf}, with the parameters listed in their Table 3. We fix the normalization of the resulting $f_+^{}(\hat s)$ and $f_0^{}(\hat s)$ such that using them in the branching fraction of $\tau^-\to\nu\pi^-\,\overline{\!K}{}^0$\, in the SM~\cite{Pich:2013lsa} yields the central value, \,0.808\%,\, of the corresponding Belle result~\cite{Epifanov:2007rf}.\medskip}
Thus, after integrating the differential rate over \,$(m_{\pi^+}+m_{K^+})^2\le\hat s\le(m_\tau-m_\ell)^2$,\, we obtain for~\,$\ell=e,\mu$\,
\begin{align} \label{Gt2lpiK}
{\cal B}(\tau^-\to e^-\pi^-K^+) & \,= \begin{array}[t]{l} \!\! \Big[
7.2\, \Big( \big|\texttt V_{e\tau}^{}\big|\raisebox{2pt}{$^2$}
+ \big|\texttt A_{e\tau}^{}\big|\raisebox{2pt}{$^2$} \Big)
+ 8.5\, \Big( \big|\texttt S_{e\tau}^{}\big|\raisebox{2pt}{$^2$} + \big|\texttt P_{e\tau}^{}\big|\raisebox{2pt}{$^2$} \Big)
\vspace{2pt} \\ +\; \displaystyle
2.7\, {\rm Re} \Big( \texttt A_{e\tau\,}^* \texttt P_{e\tau}^{}
- \texttt V_{e\tau\,}^* \texttt S_{e\tau}^{} \Big) \Big]
\frac{10^7\rm\,GeV^4}{\Lambda_{\textsc{np}}^4} \,, \end{array}
\nonumber \\
{\cal B}(\tau^-\to\mu^-\pi^-K^+) & \,= \begin{array}[t]{l} \!\! \Big[
6.5\, \big|\texttt V_{\mu\tau}\big|\raisebox{2pt}{$^2$}
+ 7.7\, \big|\texttt A_{\mu\tau}\big|\raisebox{2pt}{$^2$}
+ 10\, \big|\texttt S_{\mu\tau}\big|\raisebox{2pt}{$^2$}
+ 6.6\, \big|\texttt P_{\mu\tau}\big|\raisebox{2pt}{$^2$}
\vspace{2pt} \\ +\; \displaystyle
2.3\, {\rm Re} \Big( \texttt A_{\mu\tau\,}^* \texttt P_{\mu\tau}^{} \Big)
- 3.0\, {\rm Re} \Big( \texttt V_{\mu\tau\,}^* \texttt S_{\mu\tau}^{} \Big) \Big]
\frac{10^7\rm\,GeV^4}{\Lambda_{\textsc{np}}^4} \,. \end{array}
\end{align}
As already mentioned, ${\cal B}(\tau^-\to\ell^-\pi^+K^-)$ is equal to ${\cal B}(\tau^-\to\ell^-\pi^-K^+)$ except the subscript $\ell\tau$ of the couplings is replaced with $\tau\ell$ and a minus sign is added to $\texttt S_{\tau\ell}$.

From Eqs.\,\,(\ref{Btau2eKS})-(\ref{Gt2lpiK}), we notice that these modes do not all probe the same set
of couplings and hence are complementary in their sensitivity to the NP contributions.
The existing experimental limits [at 90\% confidence level (CL)] for these decays are \cite{Tanabashi:2018oca}
\begin{align} \label{lKlimits}
& {\cal B}(\tau^-\to e^-K_S) ~<\, 2.6\times10^{-8} \,, &
& {\cal B}(\tau^-\to\mu^-K_S) ~<\, 2.3\times10^{-8} \,,
\end{align}
\begin{align} \label{lK*limits}
& {\cal B}(\tau^-\to  e^-K^{*0}) \,<\, 3.2\times10^{-8} \,, &
& {\cal B}(\tau^-\to\mu^-K^{*0}) \,<\, 5.9\times10^{-8} \,,
\nonumber \\
& {\cal B}(\tau^-\to  e^-\,\overline{\!K}{}^{*0}) \,<\, 3.4\times10^{-8} \,, &
& {\cal B}(\tau^-\to\mu^-\,\overline{\!K}{}^{*0}) \,<\, 7.0\times10^{-8} \,,
\end{align}
\begin{align} \label{lpiKlimits}
& {\cal B}(\tau^-\to  e^-\pi^-K^+) \,<\, 3.1\times10^{-8} \,, &
& {\cal B}(\tau^-\to\mu^-\pi^-K^+) \,<\, 4.5\times10^{-8} \,,
\nonumber \\
& {\cal B}(\tau^-\to  e^-\pi^+K^-) \,<\, 3.7\times10^{-8} \,, &
& {\cal B}(\tau^-\to\mu^-\pi^+K^-) \,<\, 8.6\times10^{-8} \,.
\end{align}
We note that the \,$\tau\to\ell K^{*0},\ell\,\overline{\!K}{}^{*0}$\, numbers in Eq.\,(\ref{lK*limits}) come from searches by Belle \cite{Nishio:2008zx,Miyazaki:2011xe} which selected $K^{*0}$ and $\,\overline{\!K}{}^{*0}$ candidates from $\pi^\mp K^\pm$ pairs with invariant-masses around the $K^*(892)$ mass.
Since, on the other hand, the $\pi K$ pair in each of the \,$\tau\to\ell\pi^\mp K^\pm$\, modes proceeds in principle from all possible resonant contributions $[K^*(892)$ and its heavier counterparts] as well as nonresonant ones, the limits in Eq.\,(\ref{lpiKlimits}), from a separate Belle search~\cite{Miyazaki:2012mx}, can reasonably be assumed to have no correlation with the \,$\tau\to\ell K^{*0},\ell\,\overline{\!K}{}^{*0}$\, ones.

\subsection{\bf Other modes\label{others'}}

The \,$K\to\pi\nu\bar\nu$\, modes are sensitive to $c_{1,2,4}^{\ell\tau,\tau\ell}$ according to Eqs.\,(\ref{K2pnn})-(\ref{Wlt}).
In view of the SM predictions \,${\mathcal B}(K^+\to\pi^+\nu\nu)=\big(8.5_{-1.2}^{+1.0}\big)\times10^{-11}$\, and \,${\mathcal B}(K_L\to\pi^0\nu\bar\nu)=\big(3.2_{-0.7}^{+1.1}\big)\times10^{-11}$ \cite{Bobeth:2017ecx} and the data \,${\mathcal B}(K^+\to\pi^+\nu\nu)=1.7(1.1)\times10^{-10}$ \cite{Tanabashi:2018oca} and \,${\mathcal B}\big(K_L\to\pi^0\nu\bar\nu\big)<3.0\times10^{-9}$\, at 90\% CL~\cite{Ahn:2018mvc}, we  impose \,$\Delta{\cal B}_{K^+}<2.7\times10^{-10}$\, and
\,$\Delta{\cal B}_{K_L}<3.0\times10^{-9}$\, at 90\% CL on the NP contributions in Eq.\,(\ref{K2pnn}).
The $\Delta{\cal B}_{K^+}$ bound is clearly stricter than the $\Delta{\cal B}_{K_L}$ one and translates into \,$|W_{\ell\tau,\tau\ell}|<3.9$.\, Improvement on this bound from NA62 is expected in the near future.

The \,$|\Delta S|=0$\, decays \,$\tau\to\ell\pi^0,\ell\rho^0$\, can potentially probe $c_{1,2,5}^{\ell\tau,\tau\ell}$ as Eq.\,(\ref{lpi}) indicates.
The needed decay constants are \,$f_\pi^{}=130.2(1.7)$\,MeV \cite{Tanabashi:2018oca} and \,$f_\rho^{}=210.5(4)$\,MeV,\, the latter having been extracted from the data on \,$\tau^-\to\nu\rho^-$\, assuming isospin symmetry and no NP in this channel.\footnote{We have employed \,${\cal B}(\tau^-\to\nu\rho^-)=0.2549(9)$\, and \,$m_{\rho^-}=775.11$ MeV\, from \cite{Tanabashi:2018oca}.\medskip}  We then get
\begin{align} \label{Btau2epi}
{\cal B}(\tau^-\to e^-\pi^0) & \,=\, 2.5\, \Big( \big|c_1^{e\tau}-c_2^{e\tau}\big| \raisebox{1pt}{$^2$} + \big|c_5^{e\tau}\big| \raisebox{1pt}{$^2$} \Big) \frac{10^6\rm\,GeV^4}{\Lambda_{\textsc{np}}^4} \,,
\nonumber \\
{\cal B}(\tau^-\to\mu^-\pi^0) & \,=\, 2.4\, \Big( \big|c_1^{\mu\tau}-c_2^{\mu\tau}\big| \raisebox{1pt}{$^2$} + \big|c_5^{\mu\tau}\big| \raisebox{1pt}{$^2$} \Big) \frac{10^6\rm\,GeV^4}{\Lambda_{\textsc{np}}^4} \,, &&&
\end{align}
\begin{align} \label{Btau2erho}
{\cal B}(\tau^-\to e^-\rho^0) & \,=\, 5.9\, \Big( \big|c_1^{e\tau}-c_2^{e\tau}\big| \raisebox{1pt}{$^2$} + \big|c_5^{e\tau}\big| \raisebox{1pt}{$^2$} \Big) \frac{10^6\rm\,GeV^4}{\Lambda_{\textsc{np}}^4} \,,
\nonumber \\
{\cal B}(\tau^-\to\mu^-\rho^0) & \,=\, \Big( 2.7\, \big|c_1^{\mu\tau}-c_2^{\mu\tau}+c_5^{\mu\tau}\big| \raisebox{1pt}{$^2$} + 3.1\, \big|c_1^{\mu\tau}-c_2^{\mu\tau}-c_5^{\mu\tau}\big| \raisebox{1pt}{$^2$} \Big) \frac{10^6\rm\,GeV^4}{\Lambda_{\textsc{np}}^4} \,.
\end{align}
The numerical factors in Eqs.\,\,(\ref{Btau2epi}) and (\ref{Btau2erho}) are smaller than those of the \,$|\Delta S|=1$\, modes in Sect.\,\ref{tau2lds} partly because of the aforementioned CKM suppression factor, \,$|V_{ud}V_{us}|^2\simeq0.048$.\,
Comparing Eqs.\,\,(\ref{Btau2epi}) and (\ref{Btau2erho}) to the existing data at 90\% CL \cite{Tanabashi:2018oca}
\begin{align}
{\cal B}(\tau^-\to e^-\pi^0) & \,<\, 8.0\times10^{-8} \,, &
{\cal B}(\tau^-\to\mu^-\pi^0) & \,<\, 1.1\times10^{-7} \,, &
\nonumber \\
{\cal B}(\tau^-\to e^-\rho^0) & \,<\, 1.8\times10^{-8} \,, &
{\cal B}(\tau^-\to\mu^-\rho^0) & \,<\, 1.2\times10^{-7} \,, &
\end{align}
we see that limits on $c_{1,2,5}^{\ell\tau,\tau\ell}$ from \,$\tau^\pm\to\ell^\pm\rho^0$\, are stronger than from \,$\tau^\pm\to\ell^\pm\pi^0$.\,
However, at present they are not competitive to \,$K\to\pi\nu\bar\nu$\, and the \,$|\Delta S|=1$\, $\tau$-decays in bounding these coefficients.
Neither are other \,$|\Delta S|=0$\, semileptonic channels, such as \,$\tau\to\ell(\eta,\omega,\pi^+\pi^-)$.

The same coefficients contribute to \,$J/\psi\to\ell^\pm\tau^\mp$.\,
Using \,$f_{J/\psi}^{}=407(5)$\,MeV\, extracted from the measured \,$J/\psi\to e^+e^-$\, rate~\cite{Tanabashi:2018oca}, we find
\begin{align}
{\cal B}(J/\psi\to\ell^-\tau^+) & \,=\, 4.4\, \Big( \big|c_1^{\ell\tau}-c_2^{\ell\tau}\big|\raisebox{2pt}{$^2$}
+ \big|c_5^{\ell\tau}\big|\raisebox{2pt}{$^2$} \Big)
\frac{\rm GeV^4}{\Lambda_{\textsc{np}}^4}
\end{align}
from Eq.\,(\ref{GJ2lt}) and the same expression for ${\cal B}(J/\psi\to\ell^+\tau^-)$
but with $c_k^{\ell\tau}$ replaced by $c_k^{\tau\ell}$.
From direct searches, \,${\cal B}(J/\psi\to e^\pm\tau^\mp)<8.3\times10^{-6}$\, and
\,${\cal B}(J/\psi\to\mu^\pm\tau^\mp)<2.0\times10^{-6}$\, both at 90\% CL~\cite{Tanabashi:2018oca}.
It follows that these modes are far less sensitive to $c_{1,2,5}^{\ell\tau,\tau\ell}$ than all the $\tau$ and $K$ decays discussed above, although future quests for \,$J/\psi\to\ell^\pm\tau^\mp$\, by BESIII may improve upon the current branching-fraction limits by up to two orders of magnitude~\cite{Xiaoshen:2018tqn}.

Unlike the other decays addressed in this subsection, \,${\cal P}^+\to e^+\nu$\, for \,${\cal P}=\pi,K,D,D_s$\, can probe the (pseudo)scalar couplings $c_6^{\ell\tau,\tau\ell}$ and $c_{6\prime}^{\ell\tau,\tau\ell}$ according to Eqs.\,(\ref{DGM2enu})-(\ref{CM}).
The empirical limits, at 90\% CL, on NP effects in these modes are
\begin{align} \label{P2en}
{\cal B}(\pi^+\to e^+\nu ) & \,<\, 6.6\times 10^{-7} \,, &
{\cal B}(K^+\to e^+\nu) & \,<\, 1.2\times 10^{-7} \,, \nonumber \\
{\cal B}(D^+\to e^+\nu) & \,<\, 8.8\times 10^{-6} \,, &
{\cal B}(D_s^+\to e^+\nu) & \,<\, 8.3\times 10^{-5} \,,
\end{align}
where the numbers in the first line correspond to the 90\%-CL ranges of the errors in the observed values \,${\cal B}(\pi^+\to e^+\nu )=(1.230\pm0.004)\times10^{-4}$\, and \,${\cal B}(K^+\to e^+\nu)=(1.582\pm0.007)\times10^{-5}$ \cite{Tanabashi:2018oca}.
For numerical comparison with Eq.\,(\ref{P2en}), we adopt \,$f_D^{}=211.9$\,MeV\, and \,$f_{D_s}^{}=249$\,MeV\,
\cite{Tanabashi:2018oca} besides the $f_{\pi,K}^{}$ numbers quoted earlier.

\begin{table}[t!]
\center{\begin{tabular}{|c|c||c||c|}
     \hline
$k$ & $\begin{array}{c}\rm Lepton~flavor \vspace{-2pt} \\ \rm indices~\,(\textsl{\texttt f}_1\textsl{\texttt f}_2) \end{array}$ &
Upper bound on \,$\big|c_k^{\textsl{\texttt f}_1\textsl{\texttt f}_2}\big|\displaystyle \bigg(\frac{1\rm~TeV}{\Lambda_{\textsc{np}}}\bigg)\!\!\raisebox{9pt}{$^2$}$ & Process
\\ \hline \hline
$1,2$, or 4 & $e\tau,\mu\tau,\tau e$, or $\tau\mu$ & $4.0\times10^{-4}$ & $K^+\to\pi^+\nu\bar\nu$
\\ \hline
3 or 5 &   $e\tau$ & 0.012 & $\tau^-\to  e^-K^{*0}$
\\ \hline
3 or 5 & $\tau e$  & 0.012 & $\tau^-\to  e^-\,\overline{\!K}{}^{*0}$
\\ \hline
3 or 5 & $\mu\tau$ & 0.017 & $\tau^-\to\mu^-K^{*0}$
\\ \hline
3 or 5 & $\tau\mu$ & 0.018 & $\tau^-\to\mu^-\,\overline{\!K}{}^{*0}$
\\ \hline
6 &   $e\tau$ & 0.014  & $\tau^-\to e^-\pi^-K^+$
\\ \hline
6         &  $\tau e$  & $1.9\times10^{-3}$ & $\pi^+\to e^+\nu$
\\ \hline
6$\prime$ &   $e\tau$  & $1.3\times10^{-4}$ & $K^+\to e^+\nu$
\\ \hline
6$\prime$ &  $\tau e$  & 0.015   & $\tau^-\to e^-\pi^+K^-$
\\ \hline
6 or 6$\prime$ & $\mu\tau$ & 0.014 & $\tau^-\to\mu^-K_S$
\\ \hline
6 or 6$\prime$ & $\tau\mu$ & 0.023 & $\tau^-\to\mu^-\pi^+K^-$
\\ \hline \hline
\end{tabular}
\caption{The strongest upper-bound on each of the coefficients $c_k^{\textsl{\texttt f}_1\textsl{\texttt f}_2}$ if only one of them is nonzero at a~time and the processes which provide the constraints. Note that the lepton-flavor index ($\textsl{\texttt f}_1$ or $\textsl{\texttt f}_2$) can be carried by the neutrino.} \label{1coupling}} \medskip
\end{table}

\subsection{\bf Constraints on \boldmath$c_k^{\ell\tau,\tau\ell}$\label{constr}}

We entertain the possibility that only one of the coefficients $c_k^{\ell\tau,\tau\ell}$ is nonzero at a time.
In this case, after comparing the calculated branching fractions and their experimental data described in the preceding two subsections, we collect in Table \ref{1coupling} the best upper-bound, and the process supplying the corresponding constraint, on each coefficient.
Evidently, \,$K^+\to\pi^+\nu\bar\nu$\, and \,$\pi^+,K^+\to e^+\nu$\, produce the strongest restrictions to date on a number of these couplings.
If NA62 reaches its goal of testing the SM prediction with 10\% precision \cite{Lurkin:2018gdo}, the bound from \,$K^+\to\pi^+\nu\bar\nu$\, in Table\,\ref{1coupling} will be improved by roughly a factor of 4.
The bounds derived from lepton-flavor-violating $\tau$ decays may be lowered as much as 10 times by Belle II, which aims at reducing their branching-fraction limits by 2 orders of magnitude with its expected full dataset~\cite{Kou:2018nap}.

\section{Conclusions\label{conclusions}}

We have outlined the existing constraints on LFV in \,$|\Delta S|=1$\, semileptonic $\tau$ decays. To do this, we first parametrized the NP responsible for LFV with all the dimension-six operators in the effective Lagrangian that respects the gauge symmetry of the SM and is appropriate for an elementary Higgs. We subsequently computed all the \,$|\Delta S|=1$\, semileptonic $\tau$ decays
with an electron or muon plus one or two mesons in the final state that have been searched for.
We finally extracted the constraints on the parameters of the effective Lagrangian using the current 90\%-CL upper limits on their respective rates.

Noticing that the gauge symmetry of the SM relates these $\tau$ decay modes to other processes, we then studied those other modes. We found that the golden rare kaon decay \,$K^+\to\pi^+\nu\bar\nu$\, places the most stringent constraint available on several of the NP couplings and that this can be further improved by the expected NA62 results in the near future. Moreover, the measured \,$\pi^+\to e^+\nu$\, and \,$K^+\to e^+\nu$\, rates imply the strictest limits to date on a couple other of the NP couplings. Our numerical findings are summarized in Table~\ref{1coupling}.

\section*{Acknowledgements}

The work of X.G.H. was supported in part by the NSFC (Grant Nos. 11575115 and 11735010), by Key Laboratory for Particle Physics, Astrophysics and Cosmology, Ministry of Education, and Shanghai Key Laboratory for Particle Physics and Cosmology (Grant No. 15DZ2272100), and in part by the MOST (Grant No. MOST 106-2112-M-002-003-MY3). 
We thank Hai-Bo Li for information on experimental matters.

\appendix

\section{Feynman Rules}\label{feynrules}

\begin{table}[b!] \smallskip
\center{
\begin{tabular}{|c|c|}
     \hline
$\begin{array}[c]{c}\rm Flavor\vspace{-5pt}\\\rm structure\end{array}$ & Feynman rule
\\ \hline \hline
$(\bar d s) \big(\bar\ell\tau\big)$ &
$\big(c_1^{\ell\tau}+c_2^{\ell\tau}\big) L_\eta^{} ${\footnotesize$\otimes$}$ L^\eta
+ c_3^{\ell\tau} R_\eta^{} ${\footnotesize$\otimes$}$ R^\eta
+ c_4^{\ell\tau} R_\eta^{} ${\footnotesize$\otimes$}$ L^\eta
+ c_5^{\ell\tau} L_\eta^{} ${\footnotesize$\otimes$}$ R^\eta
+ c_6^{\ell\tau} \tilde{\textsl{\texttt L}} \otimes \tilde{\textsl{\texttt R}}
+ c_{6\prime}^{\ell\tau} \tilde{\textsl{\texttt R}} \otimes \tilde{\textsl{\texttt L}}$
\\  \hline
$(\bar d s) (\bar\nu_\ell\nu_\tau )$ &
$\big(c_1^{\ell\tau}-c_2^{\ell\tau}\big) L_\eta^{} ${\footnotesize$\otimes$}$ L^\eta
      + c_4^{\ell\tau} R_\eta^{} ${\footnotesize$\otimes$}$ L^\eta$
\\ \hline
$(\bar u u) \big(\bar\ell\tau\big)$ &
$V_{ud}^{}V_{us\,}^* \big[ \big(c_1^{\ell\tau} - c_2^{\ell\tau}\big) L_\eta^{} $$\otimes$$ L^\eta
+ c_5^{\ell\tau} L_\eta^{} $$\otimes$$ R^\eta \big]$
\\ \hline
$(\bar u c) \big(\bar\ell\tau\big)$ &
$V_{ud}^{}V_{cs\,}^* \big[ \big(c_1^{\ell\tau}-c_2^{\ell\tau}\big) L_\eta^{} ${\footnotesize$\otimes$}$ L^\eta + c_5^{\ell\tau} L_\eta^{} ${\footnotesize$\otimes$}$ R^\eta \big]$
\\  \hline
$(\bar c u) \big(\bar\ell\tau\big)$ &
$V_{cd}^{}V_{us\,}^* \big[ \big(c_1^{\ell\tau}-c_2^{\ell\tau}\big) L_\eta^{} ${\footnotesize$\otimes$}$ L^\eta + c_5^{\ell\tau} L_\eta^{} ${\footnotesize$\otimes$}$ R^\eta \big]$
\\  \hline
$(\bar c c) \big(\bar\ell\tau\big)$ &
$V_{cd}^{}V_{cs\,}^* \big[ \big(c_1^{\ell\tau}-c_2^{\ell\tau}\big) L_\eta^{} ${\footnotesize$\otimes$}$ L^\eta + c_5^{\ell\tau} L_\eta^{} ${\footnotesize$\otimes$}$ R^\eta \big]$
\\  \hline
$(\bar u u) (\bar\nu_\ell\nu_\tau)$ &
$V_{ud}^{}V_{us\,}^* \big(c_1^{\ell\tau}+c_2^{\ell\tau}\big) L_\eta^{} ${\footnotesize$\otimes$}$ L^\eta$
\\  \hline
$(\bar u c) (\bar\nu_\ell\nu_\tau)$ &
$V_{ud}^{}V_{cs\,}^* \big(c_1^{\ell\tau}+c_2^{\ell\tau}\big) L_\eta^{} ${\footnotesize$\otimes$}$ L^\eta$
\\  \hline
$(\bar c u) (\bar\nu_\ell\nu_\tau)$ &
$V_{cd}^{}V_{us\,}^* \big(c_1^{\ell\tau}+c_2^{\ell\tau}\big) L_\eta^{} ${\footnotesize$\otimes$}$ L^\eta$
\\  \hline
$(\bar c c) (\bar\nu_\ell\nu_\tau)$ &
$V_{cd}^{}V_{cs\,}^* \big(c_1^{\ell\tau}+c_2^{\ell\tau}\big) L_\eta^{} ${\footnotesize$\otimes$}$ L^\eta$
\\  \hline
$(\bar d u) (\bar\nu_\ell\tau)$ & $V_{us\,}^* \big( 2 c_2^{\ell\tau} L_\eta^{} ${\footnotesize$\otimes$}$ L^\eta + c_6^{\ell\tau}\, \tilde{\textsl{\texttt L}} \otimes \tilde{\textsl{\texttt R}} \big)$
\\  \hline
$(\bar d c) (\bar\nu_\ell\tau)$
      & $V_{cs\,}^* \big( 2c_2^{\ell\tau} L_\eta^{} ${\footnotesize$\otimes$}$ L^\eta
      + c_6^{\ell\tau}\, \tilde{\textsl{\texttt L}} \otimes \tilde{\textsl{\texttt R}} \big)$
\\  \hline
 $ (\bar u s) \big(\bar\ell\nu_\tau\big)$ & $V_{ud}^{}\, \big( 2 c_2^{\ell\tau} L_\eta^{} ${\footnotesize$\otimes$}$ L^\eta + c_{6\prime}^{\ell\tau}\, \tilde{\textsl{\texttt R}} \otimes \tilde{\textsl{\texttt L}} \big)$
\\  \hline
 $ (\bar c s) \big(\bar\ell\nu_\tau\big)$ & $V_{cd}^{}\, \big( 2 c_2^{\ell\tau} L_\eta^{} ${\footnotesize$\otimes$}$ L^\eta + c_{6\prime}^{\ell\tau}\, \tilde{\textsl{\texttt R}} \otimes \tilde{\textsl{\texttt L}} \big)$
\\ \hline
\end{tabular}
\caption{Feynman rules from ${\cal L}_{\textsc{np}}$ in Eq.\,(\ref{Lnp}).
In the second column, each entry is to be furnished with an overall factor $\Lambda_{\textsc{np}}^{-2}$ and with the spinors of the fermions in the first column, \,$V_{\textsl{\texttt U}_i\textsl{\texttt D}_j}=({\cal V}_{\textsc{ckm}})_{ij}^{}$\, from Eq.\,(\ref{ql}), and we have defined \,$L_\eta^{}=\gamma_\eta^{}P_L^{}$,\, $R_\eta^{}=\gamma_\eta^{}P_R^{}$,\, $\tilde{\textsl{\texttt L}}=P_L^{}$,\, and \,$\tilde{\textsl{\texttt R}}=P_R^{}$.\,
Since the neutrinos are nearly massless and not detected in decays, we display their weak eigenstates \,$\nu_{\textsl{\texttt E}_i}^{}=\raisebox{1pt}{\footnotesize$\sum$}_j({\cal U}_{\textsc{pmns}})_{ij}^{}\nu_j^{}$\, in the first column.} \label{tabFR}
} \vspace{-1ex}
\end{table}

The various four-fermion couplings with (2quark)(2lepton) flavor structures arising from the operators $Q_k^{\ell\tau}$ in Eq.\,(\ref{Lnp}) with \,$\ell=e,\mu$\, are collected in Table~\ref{tabFR}.
Those with the lepton flavors interchanged, (2quark)$(\bar\tau\ell)$ and (2quark)$(\bar\nu_\tau\nu_\ell)$, are readily obtainable from the corresponding entries in the table by making the change \,$c_k^{\ell\tau}\to c_k^{\tau\ell}$.\,
The Hermitian conjugates of all these couplings are additional ones with the quarks interchanged.

\bibliography{bibliotaulfv2}

\end{document}